\documentclass[a4paper,11pt]{article}
\pdfoutput=1
\usepackage{jheppub}
\usepackage{comment}
\usepackage{epstopdf}

\preprint{PITT-PACC 1312}

\usepackage{graphicx}
\usepackage{multirow}
\usepackage{subfigure}
\usepackage{soul}
\newcommand{\lsim}{\mathrel{\mathop{\kern 0pt \rlap
  {\raise.2ex\hbox{$<$}}}
  \lower.9ex\hbox{\kern-.190em $\sim$}}}
\newcommand{\gsim}{\mathrel{\mathop{\kern 0pt \rlap
  {\raise.2ex\hbox{$>$}}}
  \lower.9ex\hbox{\kern-.190em $\sim$}}}
\newcommand{\gev}{{\,{\rm GeV}}}
\newcommand{\tev}{{\,{\rm TeV}}}
\newcommand{\beq}{\begin{equation}}
\newcommand{\eeq}{\end{equation}}
\newcommand{\bea}{\begin{eqnarray}}
\newcommand{\eea}{\end{eqnarray}}

\def\ww{W W^*}
\def\zz{ZZ^*}
\def\ee{e^+ e^-}
\def\fbi{{\rm fb}^{-1}}

\def\br{{\rm Br}}
\def\cm2s{{\rm cm^{-2} s^{-1}}}

\begin{document}

\title{Potential Precision on Higgs Couplings and Total Width at the ILC}
\bigskip
\author[a]{Tao Han}
\author[a]{,~Zhen Liu}
\author[a]{and Josh Sayre}
\affiliation[a]{Pittsburgh Particle physics, Astrophysics and Cosmology Center\\
Dept.~of Physics $\&$ Astronomy,
Univ.~of Pittsburgh, 3941 O'Hara St., Pittsburgh, PA 15260, USA}
\abstract{
We outline a systematic approach to the determination of the Standard Model-like Higgs boson total width and measurable coupling parameters in a model-independent manner at the International Linear Collider (ILC) and illustrate the complementarity for operating the ILC at $250~\gev$ near the $Zh$ threshold and at $500~\gev$ and $1~\tev$ utilizing the $WW, ZZ$ fusion processes.
We perform detailed simulations for an important contributing channel to the coupling determination and for invisible decays. Without model assumptions, and combining the information for the coupling ratios from the LHC, the total width can be determined to an accuracy of about $6\%$, and the couplings for the observable channels can be measured to the $(3-5)\%$ level at $250~\gev$, reaching $(1-3)\%$ level including the $500~\gev$ results, with further improvements possible with a $1~\tev$ run.
The best precision for the branching fraction measurement of the Higgs to invisible modes can be reached at $0.5-0.7\%$ around the $Zh$ threshold.
Further studies from $ZZ$ fusion at higher energies may provide significant improvement for the measurements.
With modest theory assumptions, the width and coupling determinations can be further improved to the percent or sub-percent level.
}

\maketitle


\section{Introduction}
The discovery of a Higgs boson at the LHC \cite{Aad:2012tfa,Chatrchyan:2012ufa} completes the simple structure of the Standard Model (SM). Yet, a profound question remains: Is this rather light, weakly-coupled boson nothing but a SM Higgs or it is a first manifestation of a deeper theory? While we look forward to a long and hopefully fruitful run at the LHC as it continues to search for direct evidence of new physics beyond the SM, it will be very important to determine the Higgs boson's properties as accurately as possible at the LHC and future collider facilities. It may be that the Higgs itself is our first access to the next regime of physics.

The key properties of the Higgs boson are the strengths of its couplings to other particles. The overall strength of these couplings, at least to particles lighter than the Higgs itself, is characterized by the total width
$\Gamma_h$.
Because of the broad spread of the partonic energy distribution, limited energy-momentum resolution for final state particles,  and the large SM backgrounds in the LHC environment,
there is essentially no way to measure its total width or any partial width to a {\it desirable accuracy} without additional theoretical assumptions~\cite{Duhrssen:2004cv,Peskin:2012we,Dobrescu:2012td}. Assuming an upper limit for a Higgs coupling, such as that of $hWW$ and $hZZ$, then an upper bound for the total width can be inferred~\cite{Dobrescu:2012td}.
Around $ (10 -40)\times\Gamma_h$ accuracy can be achieved by measuring the interference effect which shifts the invariant mass distribution of the ZZ/$\gamma\gamma$ mode~\cite{Dixon:2013haa}, and by measuring $pp\to ZZ$ differential cross sections with $m_{ZZ}>300~\gev$~\cite{Caola:2013yja,Campbell:2013una}.
At a lepton collider optimized for Higgs boson studies, such as an International Linear Collider (ILC) or a circular $e^+e^-$ collider (TLEP)~\cite{Gomez-Ceballos:2013zzn},
the $ h ZZ$ coupling, and thus the partial decay width $\Gamma(h\to ZZ)$ can be measured to a good accuracy~\cite{Abe:2010aa} by measuring inclusive Higgs cross sections. The total decay width then can be indirectly determined. For review on ILC, see the ILC Technical Design Report (TDR)~\cite{Baer:2013cma}. See, e.g.,~\cite{Barger:2013ofa} for an estimate of sensitivity in a 2-Higgs Doublet Model.

In this paper, we revisit the issue of to what extent an ILC, combined with expected LHC inputs, would be able to precisely determine the Higgs total width and individual couplings. We follow a systematic approach in a model-independent manner. We discuss the leading and subleading contributions for total width and perform studies on two typical processes at the ILC. We find the optimal results for the Higgs couplings, the invisible decay mode, and total width determinations. Finally, we consider the effects of adding mildly model-dependent assumptions, which can significantly improve the precision.


\section{Formalism for Higgs Width Determination}
\label{sec:approach}

\subsection{General Approach to the Higgs Width}
\label{sec:paths}

The total width of the SM Higgs boson is predicted to be about $4.2$ MeV.
For such a narrow width, only a muon collider may provide a sufficiently small energy spread to directly measure the width~\cite{Barger:1996jm,Han:2012rb,Conway:2013lca,Alexahin:2013ojp}. At the LHC and ILC, the smearing effects from energy spread for both the initial state and final state override the Breit-Wigner resonance distribution, although it may have a visible effect in differential cross section distributions involving interference with the background diagrams \cite{Dixon:2013haa,Caola:2013yja,Campbell:2013una}.

The Narrow Width Approximation (NWA) allows us to write a total $s$-channel cross section as a production cross section multiplied by the branching fraction (Br) of the Higgs boson decay
\beq
\sigma_{AB}
\simeq \frac {\sigma(A \rightarrow h)\Gamma(h\rightarrow B)} {\Gamma_h}
\propto\frac {g_A^2 g_B^2} {\Gamma_h} ,
\label{eq:sigab}
\eeq
where we have symbolically denoted the Higgs production via $A$ with an $AAh$ coupling ($g_A$), and the subsequent decay to $B$ with a $BBh$ coupling ($g_B$).
There is a well-known ``scaling degeneracy'' of the NWA cross section, namely the cross section is invariant under the scaling of related couplings by $\kappa$ and the total width of the resonant particle by $\kappa^4$. This demonstrates the incapability of hadron colliders to determine the couplings and width in a
model-independent fashion.
With certain modest assumptions, one may obtain some bounds on the total width as recently discussed in~\cite{Barger:2012hv,Peskin:2012we,Dobrescu:2012td,Klute:2013cx,Dawson:2013bba}.

When determining the accuracies from the measurements, we take the form of Eq.~(\ref{eq:sigab}) in a general  sense. We consider all available measurements from, for instance, $AA$-fusion, $Ah$-associated production etc.\ to extract $g_A^2$. This form also allows for interchanging production and decay since $g_A^2$ and $g_B^2$ are on equal footing.

In order to break the ``scaling degeneracy'' without assumptions on the couplings, we must go beyond the simple form $\sigma_{AB}$. The most efficient process for this purpose is the inclusive Higgs production cross section from the coupling to $A$.
This can be measured when we know the information about the incoming and outgoing particles aside from the Higgs boson in the process. The best-known example is the ``Higgstrahlung'' process $e^+e^- \rightarrow hZ$, where one can construct the recoil mass by the well-measured $Z$ decay products
\beq
M_{rec}^{2} = (p_{e^+e^-} -p_Z)^2  = s + M^{2}_{Z^{*}} - 2 \sqrt s E_{Z}.
\label{eq:mrec3}
\eeq
The peak near $m_{h}$ selects the signal events and the Higgs decay modes are all accounted for inclusively. Thus the inclusive cross section $\sigma^{inc}_Z$ and equivalently $g_Z^2$ can be measured since the factor $\sum_{all} g_{B}^{2} / \Gamma_{h}$ is unity.

With the inclusive cross section $\sigma^{inc}_A$ (equivalently $g_A^2$) measured, one can readily perform the extraction to the total width $\Gamma_h$ by utilizing exclusive cross sections:\\
$\bullet$
directly measuring $\sigma_{AA}$
\beq
\Gamma_h = \frac{(g_A^2)^2}{(g_A^2 g_A^2 /\Gamma_h)} \propto g_A^2\ \frac{\sigma^{inc}_A}{\sigma_{AA}};
\label{eq:path1}
\eeq
$\bullet$
indirectly determining $\sigma_{AA}$ by inserting other cross section measurements from the ILC
\bea
\Gamma_h &=& \frac{(g_A^2)^2 (g_B^2 g_C^2 /\Gamma_h)}{(g_A^2 g_B^2/\Gamma_h) (g_A^2 g_C^2/\Gamma_h)} \propto g_A^2\ \frac{\sigma_A^{inc} \sigma_{BC}}{\sigma_{AB}\sigma_{AC}}
\label{eq:path2} \\ \nonumber
&=&
\frac{(g_A^2)^2 (g_B^2 g_C^2 /\Gamma_h)(g_D^2 g_E^2 /\Gamma_h)}{(g_A^2 g_B^2/\Gamma_h) (g_C^2 g_D^2/\Gamma_h)(g_A^2 g_E^2/\Gamma_h)}
\propto g_A^2\
\frac{\sigma^{inc}_A \sigma_{BC}\sigma_{DE}}{\sigma_{AB}\sigma_{CD}\sigma_{AE}}\\ \nonumber
&=& ... ; \nonumber
\eea
$\bullet$
more generally, indirectly determining $\sigma_{AA}$ by inserting other cross sections including those from the LHC
\beq
\Gamma_h = \frac{(g_A^2)^2}{g_A^2 g_B^2/\Gamma_h}\left(\frac {g_B^2} {g_A^2}\right)\propto g_A^2 \ \frac{\sigma^{inc}_A}{\sigma_{AB}}\left( \frac {\br_B^{}} {\br_A^{}}\right).
\label{eq:path3}
\eeq
We note that in the above method as expressed in Eqs.~(\ref{eq:path1}), (\ref{eq:path2}) and (\ref{eq:path3}), the right hand sides are fully expressed by experimental observables\footnote{We keep $g_A^2$ to make clear the transition between Eq.~(\ref{eq:sigab}) and Eq.~(\ref{eq:path1}). As stated earlier, it is a direct translation from observable $\sigma^{inc}_A$}, which can be easily and consistently used to determine precision on derived quantities such as $\Gamma_h$. In principle, the longer the  chain of measured cross-sections is in our expressions above, the more sources of uncertainties we must be concerned with, but this may allow us to utilize quantities with minimal individual uncertainties which can be advantageous.
 The most important channels to measure depend on the center of mass energy. Current plans for the ILC foresee an initial stage of running at $250$ GeV, which maximizes the Higgstrahlung cross section, and a higher-energy phase at $500$ GeV with perhaps $1$ TeV running at an upgraded machine where weak boson fusion takes over.
 This leads to rich physics interplay between combinations of contributing channels which we discuss in the following subsections Sec.~\ref{sec:250} and Sec.~\ref{sec:500}.

Determining the couplings other than $g_A^2$ from $\sigma^{inc}_A$, can be done directly once $\Gamma_h$ is known. With any exclusive cross section measurement $\sigma_{AB}$ or $\sigma_{BA}$, we can write
\bea
g_B^2\propto {\sigma_{AB}\Gamma_h \over g_A^2} .
\label{eq:gb}
\eea
However, one needs to be cautious when determining $g_B^2$ using this relation.
$\Gamma_h$ is a derived quantity and may depend heavily on $\sigma_{AB}$ as well. For proper treatment of errors, we will evaluate precision on these quantities consistently by global fitting as described in Sec.~\ref{sec:mi} and Appendix~\ref{sec:chidef}. Discussions in this subsection and following subsections clearly point out leading and sub-leading contributions and will provide guidance for current and future studies.


\subsection{The ILC at $250$ GeV}
\label{sec:250}
\subsubsection{With ILC only}

 For an electron-positron machine running near $250$ GeV the leading Higgs
production mechanism is the ``Higgstrahlung'' process $e^+e^- \rightarrow hZ$.
For $Z$ decaying to electrons and particularly to muons we can have very good resolution on recoil mass and a clear excess over expected background. Detailed simulations estimate that the inclusive cross section $\sigma(Zh)$ can be determined to a statistical uncertainty of $2.5 \%$ with $250$ fb$^{-1}$ of integrated luminosity~\cite{Li:2009fc,Li:2010wu,Li:2012taa}.

Unfortunately, $\sigma_{ZZ}$ can not be measured with great precision at the ILC due to the limited statistics from the small $Z$ leptonic branching fractions. As such, , the total width determined from Eq.~(\ref{eq:path1}) has large uncertainties. However, we can make several measurements from which an equivalent ratio of couplings to widths is derivable, as shown in Eq.~(\ref{eq:path2}).
For a Standard-Model-like Higgs boson, couplings to $b$-quarks and to $W$ and $Z$ bosons are expected to dominate, leading to high statistics for those channels. Since we are mainly interested in ratios of coupling constants and the Higgs width, $\sigma_{WZ}$ and $\sigma_{ZW}$ give us equivalent information and can potentially both be measured. To use Eq.~(\ref{eq:path2}) we must have at least one cross section that involves a Higgs coupling to non-$Z$ particles at both vertices. At the ILC this generally requires producing the Higgs via $WW$ fusion. This mechanism becomes dominant at higher energies but remains relatively small at $250$ GeV. Nonetheless, D\"{u}rig et al.~estimate that $\sigma(e^+e^- \rightarrow h\nu\nu \rightarrow b\overline{b}\nu\nu) =\sigma_{Wb}$ can be measured with $10.5 \%$ accuracy~\cite{Durig}. Then by measuring $\sigma_{Zb}$ and $\sigma_{WZ}$ one has an alternative and likely more precise determination of the Higgs width
\beq
\Gamma_h\propto g_{Z}^{2}\ \frac{\sigma^{inc}_Z\ \sigma_{Wb}} {\sigma_{ZW}\ \sigma_{Zb}}.
\label{eq:path250}
\eeq

\subsubsection{Incorporating LHC data}

The LHC Run II will accumulate a significant amount of integrated luminosity and the Higgs property will be studied to a high accuracy. Although, as discussed in Sec.~\ref{sec:paths}, it cannot resolve the inclusive Higgs measurement, we can use ratios of cross sections from the LHC in conjunction with ILC data to improve our results as described by Eq.~(\ref{eq:path3}).  The ATLAS and CMS collaborations have conducted simulations to estimate the sensitivity of
various cross section and ratio measurements with $300$ fb$^{-1}$ of data and in some cases up to $3000$ fb$^{-1}$ \cite{ATLAS:2013hta,CMS:2013xfa,ATLAS-HLnote}. In particular, with $300$ fb$^{-1}$ of data the LHC is expected to measure the Higgs decays to $\gamma\gamma$, $ZZ$, $WW$, $bb$ and $\tau\tau$ with $\sim 5-20\%$ accuracy. We can use these numbers along with measurements of $\sigma_{Zb}$, $\sigma_{Z\gamma}$, $\sigma_{ZW}$ and $\sigma_{Z\tau}$ at the ILC to determine the total width as well. Generically, either the relevant ILC cross section or the ratio coming from the LHC will have limited sensitivity so the individual combinations will have only
moderate uncertainty for the total Higgs width, but in combination with each other and the pure ILC combinations above an improved result for the width can be achieved, as will be discussed in section~\ref{sec:accur}.

\subsubsection{Invisible Decays of the Higgs}
\label{sec:hinv}
One further decay channel which it is interesting to include is the partial width for Higgs decaying to invisible particles. In the SM this is a tiny branching fraction due to $h \rightarrow ZZ \rightarrow 4\nu$ ($\br \sim 0.2 \%$). However it may be enhanced by new physics such as Higgs portal scenarios for dark-matter~\cite{Patt:2006fw}. The invisible decay cross section can be measured to high precision at the ILC. This is done by again using the recoil mass and the absence of visible final particles except for the recoiling
matter, which will be a $Z$ for our purposes at the $250$ GeV ILC. Since we only expect to measure one cross section involving the coupling to invisible particles, the invisible decay measurement does not constrain the total Higgs width in a model-independent analysis. Rather, other measurements largely fix $g_Z$ and $\Gamma_h$ which then constrain $g_{inv}$, the effective coupling to invisible final states. However, as will be discussed in Sec.~\ref{sec:md},
the invisible width can become an important constraint when applying very moderate assumptions. We have performed a fast simulation of the invisible decay sensitivity, which is detailed in Sec.~\ref{sec:inv}.

\subsection{The ILC at $500$ GeV and Beyond}
\label{sec:500}

Beyond $250$ GeV the Higgstrahlung cross section falls off and the fusion cross sections grow. At $500$ GeV $WW$ fusion is the leading process with a cross section of approximately $\sim 130$ fb. The total $Zh$ inclusive cross section is $\sim 100$ fb, however the $he^+e^-$ component is only about $3\%$ of this and similarly for the muon decay mode. At this energy, $ZZ$ fusion to $e^{+}e^{-}h$ contributes roughly twice as much cross section as $hZ$ with $Z\to e^+e^-,\mu^{+}\mu^{-}$ \cite{Gunion:1998jc}. The inclusive cross section cannot be measured as well for leptonic decays of the $Z$ but including hadronic decays it may be possible to establish $\sigma_Z^{inc}$ at $3\%$ using $500$ GeV data \cite{Asner:2013psa}. At $1$ TeV the fusion cross-sections completely dominate the production signal. For a SM-like Higgs the best individual determination of the width is expected to come from measuring $\sigma_{Wb}$,
$\sigma_{WW}$ and $\sigma_{Zb}$ with high precision. These can be put in the form of Eq.~(\ref{eq:path2})

\bea
\Gamma_h \propto g_{Z}^{2}\ \frac{\sigma^{inc}_Z \ \sigma_{Wb}^2}{\sigma_{Zb}^2\ \sigma_{WW}}.
\label{eq:width4}
\eea

Based on the statistical uncertainty expected in these channels, one finds the precision on the total width can be known with a $\sim6 \%$ error as reported in Ref.~\cite{Baer:2013cma}, and as confirmed using the numbers in Sec.~\ref{sec:mi}. This error is dominated by the uncertainty on the inclusive cross section, which is squared in our formula, and that of
the cross section $\sigma_{WW}$. Although the remaining measurements, $\sigma_{Wb}$ and $\sigma_{Zb}$ both enter quadratically, they are expected to be known to the sub-percent level and thus add only a small contribution to the total uncertainty. These uncertainties assume that the $250$ GeV run has been completed in order to obtain the best resolution on $\sigma_{Zb}$ and on $\sigma_{Zh}$. One should, however, bear in mind that these are statistical uncertainties based on SM productions and decays rates; this formula is sensitive to additional theoretical and systematic uncertainties, and to deviations from the SM.

Assuming the numbers used above, we can ask if any other channels will contribute significantly to the total model-independent width. The best candidate is one used for the $250$ GeV analysis as in Eq.~(\ref{eq:path250}).

This derivation depends linearly on the sub-percent cross sections noted above and is therefore less sensitive to any additional sources or error not included in the purely statistical determinations currently in use. It makes use of $\sigma_{WZ}$ rather than $\sigma_{WW}$. Although $\sigma_{WW}$ has a large cross section at $500$ GeV, $\sigma_{WZ}$ can be determined from
several different measurements at $250$ and $500$ GeV. We have carried out a detailed simulation of signal and background for one of these processes which we outline in Sec.~\ref{sec:500ww}.

The best constraints will come by measuring as many channels as possible, including the available information from the LHC. However, only minor improvements beyond Eq.~(\ref{eq:path2}) are possible at $500$ GeV for an approximately SM-like Higgs. The uncertainty on the inclusive cross section becomes the largest source of error and total width depends on it quadratically, as seen in Eqs.~(\ref{eq:path1}), (\ref{eq:path2}) and (\ref{eq:path3}). Due to this dependence the error on $\sigma^{inc}_Z$ contributes $5\%$ to the total width uncertainty, and improving this key measurement is crucial to any substantial improvements on the total width. As a result, we argue that inclusive measurement from $ZZ$-fusion at a 500 GeV and 1 TeV ILC deserves detailed studies for potential improvements.


\section{Simulations}
\label{sec:results}

Many analyses for specific ILC channels exist in the literature, in particular the recently published ILC TDR \cite{Baer:2013cma} and the Snowmass Report \cite{Asner:2013psa}.
We now present  two new studies in this section, that contribute to our determination of the width as motivated in sections Sec.~\ref{sec:250} and Sec.~\ref{sec:500}.

\subsection{Invisible Decays of the Higgs at $250$ GeV}
\label{sec:inv}

We perform a quick simulation of the invisible signal to estimate the sensitivity. For event generation we use the ILC-Whizard setup provided through the detector simulation package SGV3 \cite{Berggren:2012ar}. Beam profiles for several energies have been generated by GuineaPIG~\cite{Schulte:1999tx}, these include effects from Beamstrahlung and ISR. These profiles are interfaced with Whizard 1.95~\cite{Kilian:2007gr}. The output from this event generator is showered and hadronized by Pythia and the final state particles are passed to SGV, which performs a fast detector simulation. Detected particles are grouped into jets by SGV, for which we set an initially low separation cutoff. The Higgs is taken to be SM-like in its couplings with a mass of 126 GeV. We assume a beam polarization of $-0.8$ for the electron and $+0.3$ for the positron, consistent with standard assumptions used in the ILC TDR ~\cite{Baer:2013cma}.
The simulation includes generator level cuts of $M_{jj}>10~\gev$ and $|M_{ll}|>4~\gev$ where $j$ are outgoing quarks or gluons and $ll$ applies to final-final state lepton pairs and to initial-final state pairs of the same charge. To check our simulations we have performed a calculation of the $Zh \to b\overline{b}\nu\nu$ signal in our setup following the analysis of Ref.~\cite{Ono:2012ah}, which used a full detector simulation. We find good agreement on the expected number of events.

\begin{figure}[tb]
\centering
\mbox{\subfigure{
\includegraphics[angle=0,scale=0.35]{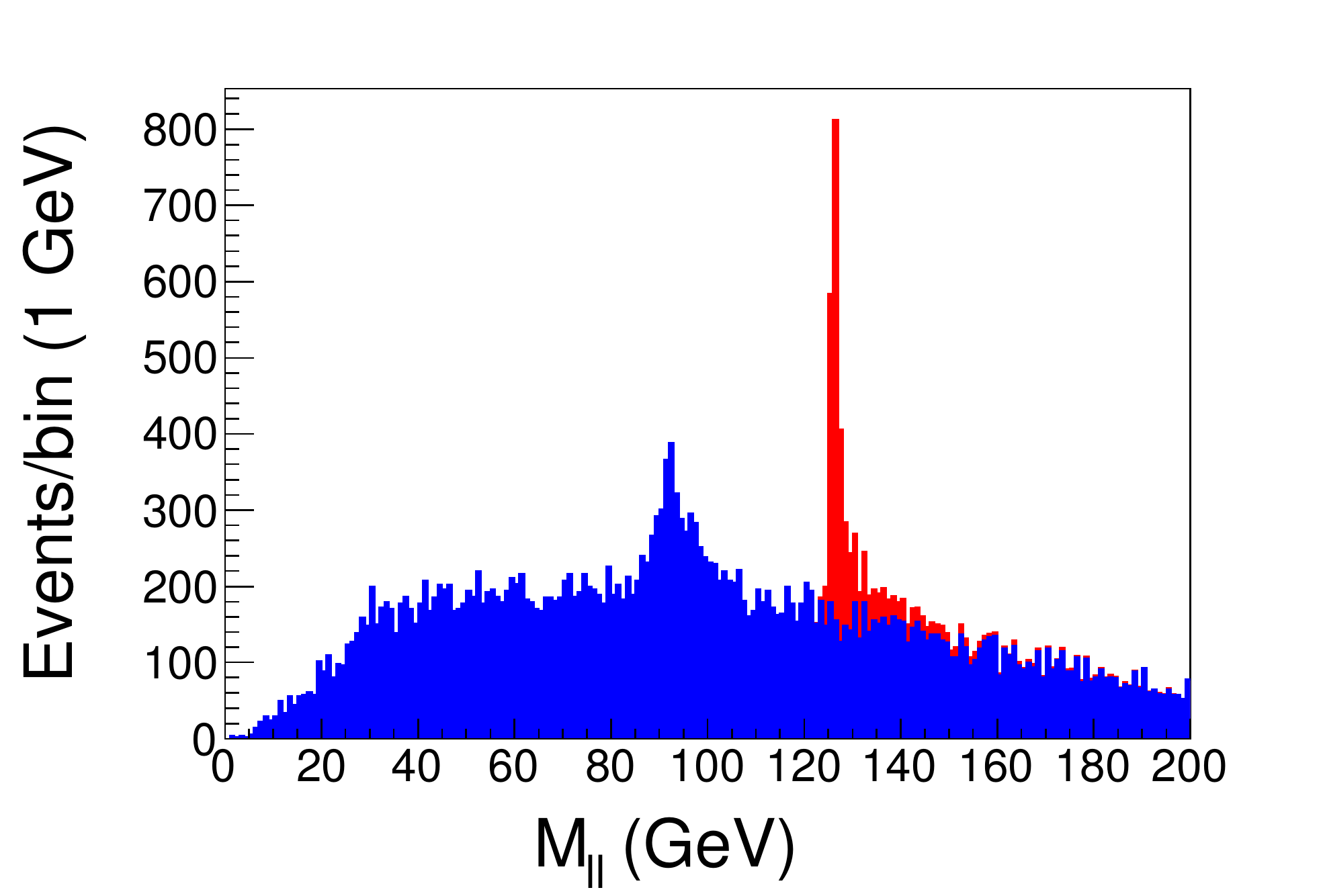}} \subfigure{
\includegraphics[angle=0,scale=0.35]{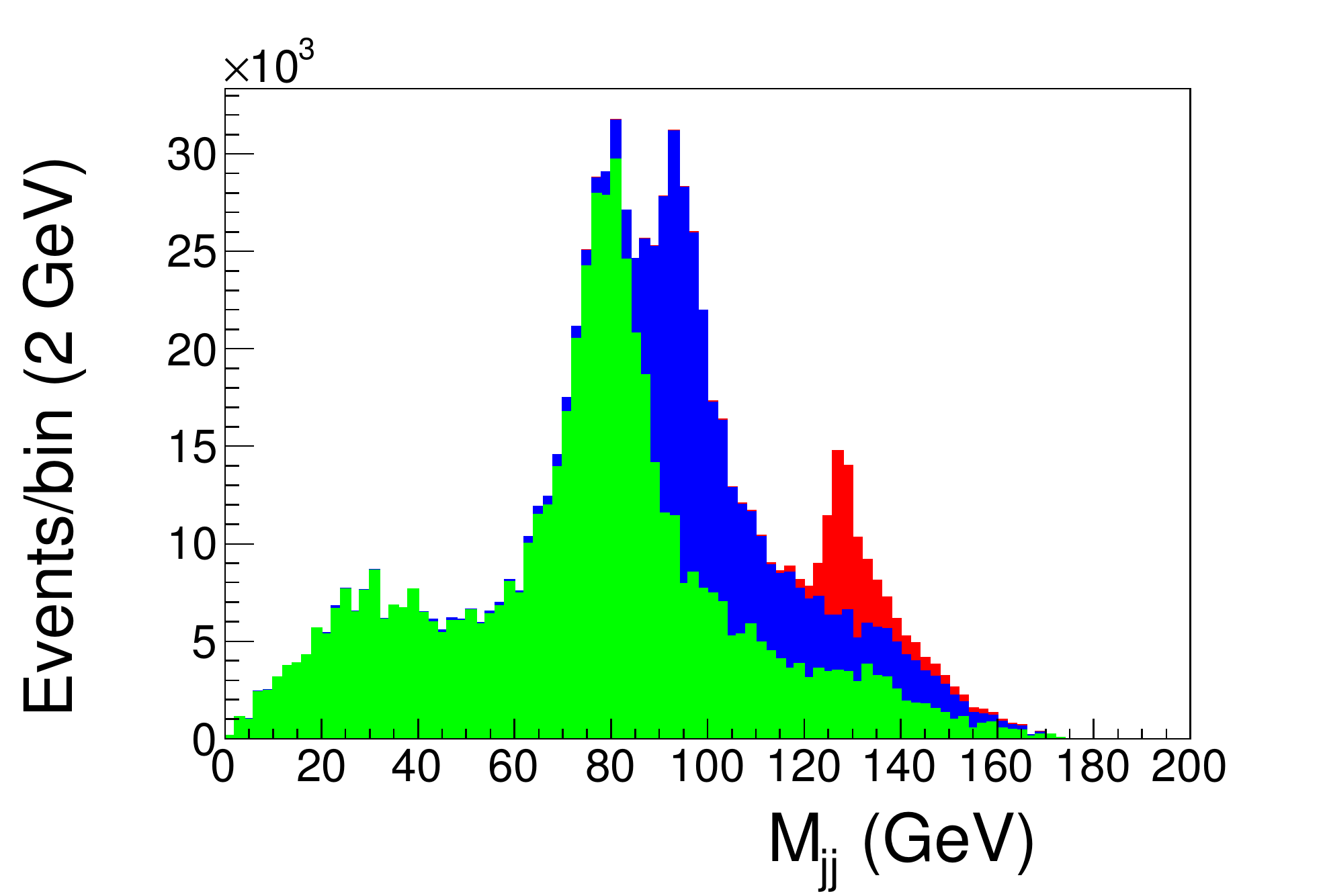}}}
\caption[]{Recoil mass distributions for the invisible decay of the Higgs boson (assuming $100 \%$ branching fraction) for $250$ fb$^{-1}$ at $250$ GeV. The Higgs signal is shown in red while the background in blue. In the left panel: recoil mass for $Z\to e^{+}e^{-}$; in the right panel: recoil mass for $Z\to jj$. The cuts are described in Sec.~\ref{sec:inv}.
}
\label{fig:invfig}
\end{figure}

For a signal sample we use the Standard Model process $e^+e^- \rightarrow Zh, h \rightarrow ZZ \rightarrow 4\nu$  as a template and scale it according to a parameterized branching fraction

\bea
\sigma_{Zh\to Z+ inv} = \sigma_{Zh}\times \br_{inv}.
\eea
We perform this analysis at the $250$ GeV energy scale and concern ourselves only with the Higgstrahlung production process. We consider two analysis channels: $Z$ decaying to leptons ($e^+e^-$ and $\mu^+\mu^-$) and $Z$ decaying to jets. The latter has lower resolution of the peak but benefits from large statistics. In Figure~\ref{fig:invfig} we present simulation results showing the recoil mass peak overlaying the major backgrounds in both channels.

We perform our analysis by imposing the following requirements: In the leptonic $Z$ case we require exactly two detected charged tracks, which are identified as either opposite sign muons or opposite sign electrons. In general the electrons are subject to bremsstrahlung radiation which degrades their momentum reconstruction. To ameliorate this we include momentum from calorimeter hits in a small region around the track when they are not associated with hadronic activity. The invariant mass of the lepton pair, $M_{ll}$ must be between $80$ and $100$ GeV. The recoil mass must satisfy $120 < M_{rec} < 150$ GeV. The magnitude of the 3-momentum of the pair must be less than $50~\gev$. Finally, the total detected momentum and total detected energy must both be within $10$ GeV of the momentum and energy of the lepton pair.

For the hadronic $Z$ decays, we first use the Durham algorithm ($k_t$ algorithm) to merge all detected particles down to two jets, which must have an invariant mass $70 <
M_{jj}<110$ GeV. The recoil mass must satisfy $120 < M_{rec} < 160$ GeV. The transverse momentum of the jet pair must be greater than $20$ GeV. To reduce the background from leptonically decaying $W$s we veto events where the highest energy charged track is greater than $35$ GeV.

The main background for the leptonic $Z$ comes from $e^+e^- \rightarrow ll\nu\nu$. In the hadronic case we see approximately equal backgrounds from $e^+e^- \rightarrow qq\nu\nu$ and $e^+e^- \rightarrow qql\nu$. We also consider
$e^+e^- \rightarrow qqll$, although it adds a negligible number of events to either channel after cuts. We present the expected number of events from signal and background in Table \ref{tab:invevnts} below. The numbers shown are for  $250$ fb$^{-1}$ of integrated luminosity and a $100 \%$ branching fraction to invisible particles for the Higgs.

\begin{table}[tb]
\centering
\begin{tabular}{|c|c|c|c|}
\hline
Process & $Z \to ee$ & $Z \to \mu\mu$ & $Z \to jj$ \\ \hline
$ee \to Zh$ & $1810$ & $1970$ & $41900$ \\ \hline
$ee \to ll\nu\nu$ & $4730$ & $3000$ & $6220$ \\ \hline
$ee \to qq\nu\nu$ & $0$ & $0$ & $20700$ \\ \hline
$ee \to qql\nu$ & $0$ & $0$ & $22600$ \\ \hline
$ee \to qqll$ & $0$ & $0$ & $84$ \\ \hline
\end{tabular}
\caption[]{Expected number of events in invisible Higgs searches for leptonically and hadronically decaying $Z$ at a $250$ GeV
ILC with $250$ fb$^{-1}$ of data.}
\label{tab:invevnts}
\end{table}

We will take the statistical uncertainty in a given channel to be $\sqrt{N_S + N_B}$ where $N_S$ and $N_B$ are the expected number of signal and background events, respectively.
Based on our numbers, given a $10 \%$ branching fraction to invisible decays, one could measure the studied cross section with a $5.4 \%$ relative accuracy. The cross section for $1\%$ branching fraction could be measured with $52\%$ relative uncertainty. A branching fraction greater than $3.5\%$ can be excluded at $95 \%$ confidence in the leptonic channel alone and as low as $0.9 \%$ can be excluded for the hadronic channel. Similar results have been reported in the Snowmass literature \cite{Asner:2013psa}.

\subsection{Estimated Sensitivity for $\sigma_{ZW}$ at $500$ GeV}
\label{sec:500ww}

To augment the sensitivity of $\sigma_{ZW}$, we carry out a Monte Carlo simulation of the signal $e^+e^- \rightarrow e^+e^-h \rightarrow e^+e^-W^+W^-$ at the ILC running at $500$ GeV.
In particular, we include signal events generated by $ZZ$ fusion graphs, which are small at $250$ GeV but comprise the majority of events at $500$ GeV.

This signal has several nice features. At $500$ GeV, after cuts, approximately two thirds of the signal is generated by $ZZ$ fusion, and one third comes from the Higgstrahlung process. We search for an on-shell Higgs decaying to one on-shell and one off-shell $W$. Each $W$ then decays either hadronically to two jets or leptonically to a charged lepton and a neutrino. We consider the all-hadronic and semi-leptonic cases for the two $W$s taken together; the all leptonic-mode makes up only a small fraction ($\sim 9 \%$) of total $WW$ decays. For both the hadronic and semi-leptonic cases the event is essentially fully reconstructible, with the neutrino momentum assumed to be equal to the missing momentum in the semi-leptonic case.

Our simulation framework is the same as described in the previous subsection. For each event, we first identify the two highest energy charged tracks which have been identified as electrons by SGV. If these are not opposite in charge sign we consider the next highest energy electron track until we find one that is of opposite sign to the highest energy track. Otherwise we discard the event. If these tracks are identified with a jet that includes seen hadronic particles, we subtract the track momentum from the jet and use the observed track momentum as the electron momentum, otherwise we identify the electron momentum with the `jet' determined from nearby calorimeter hits. 
After this process we define the number of jets with hadronic particles and energy greater than 5 GeV to be $N_{hj}$ (number of hadronic jets). We also consider potential muon tracks. Muons are not specifically identified by SGV, they appear as charged tracks that are not identified as electrons or hadronic states.
%

\begin{figure}[tb]
\centering
\mbox{\subfigure{
\includegraphics[angle=0,scale=0.35]{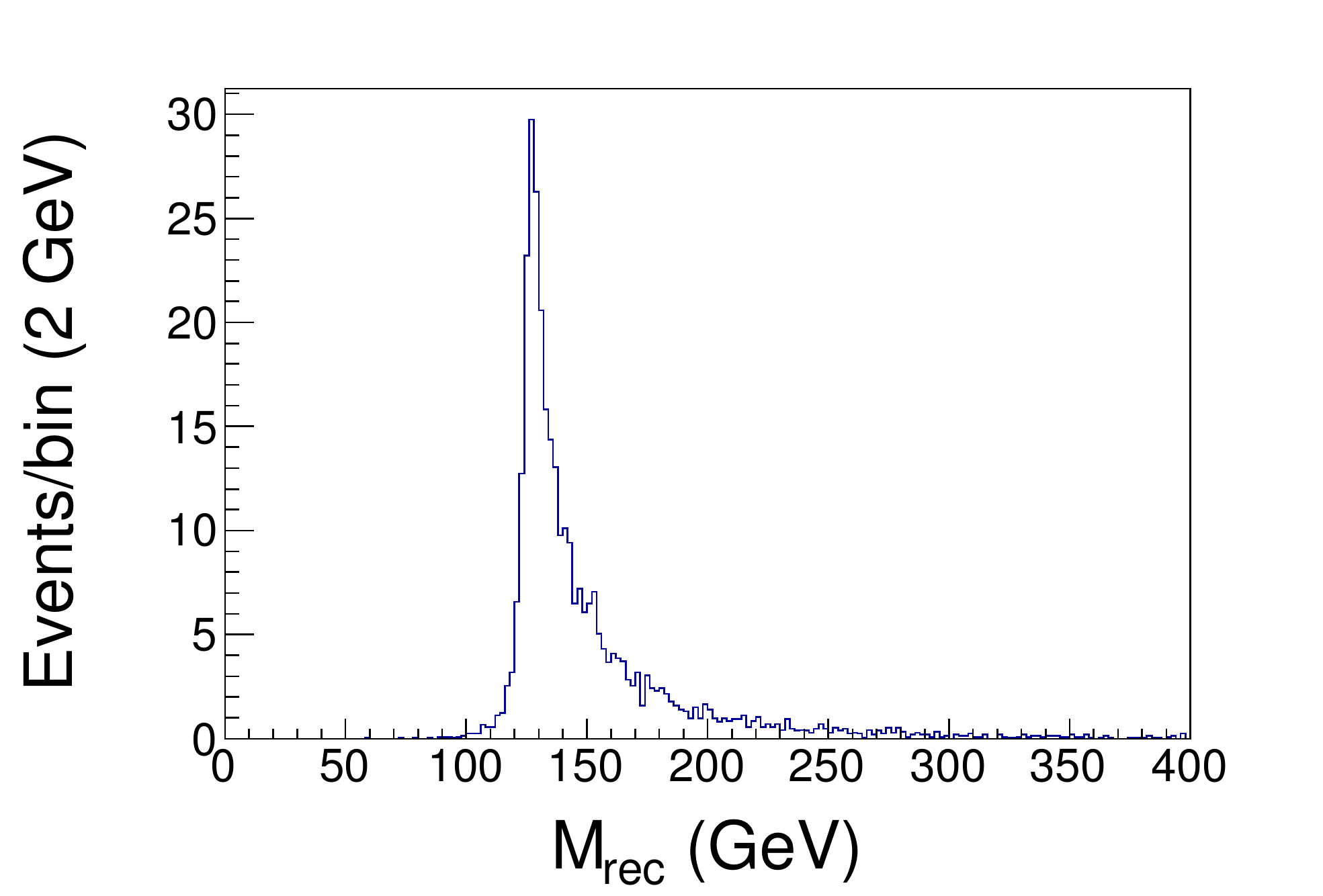}} \quad \subfigure{ \includegraphics[angle=0,scale=0.35]{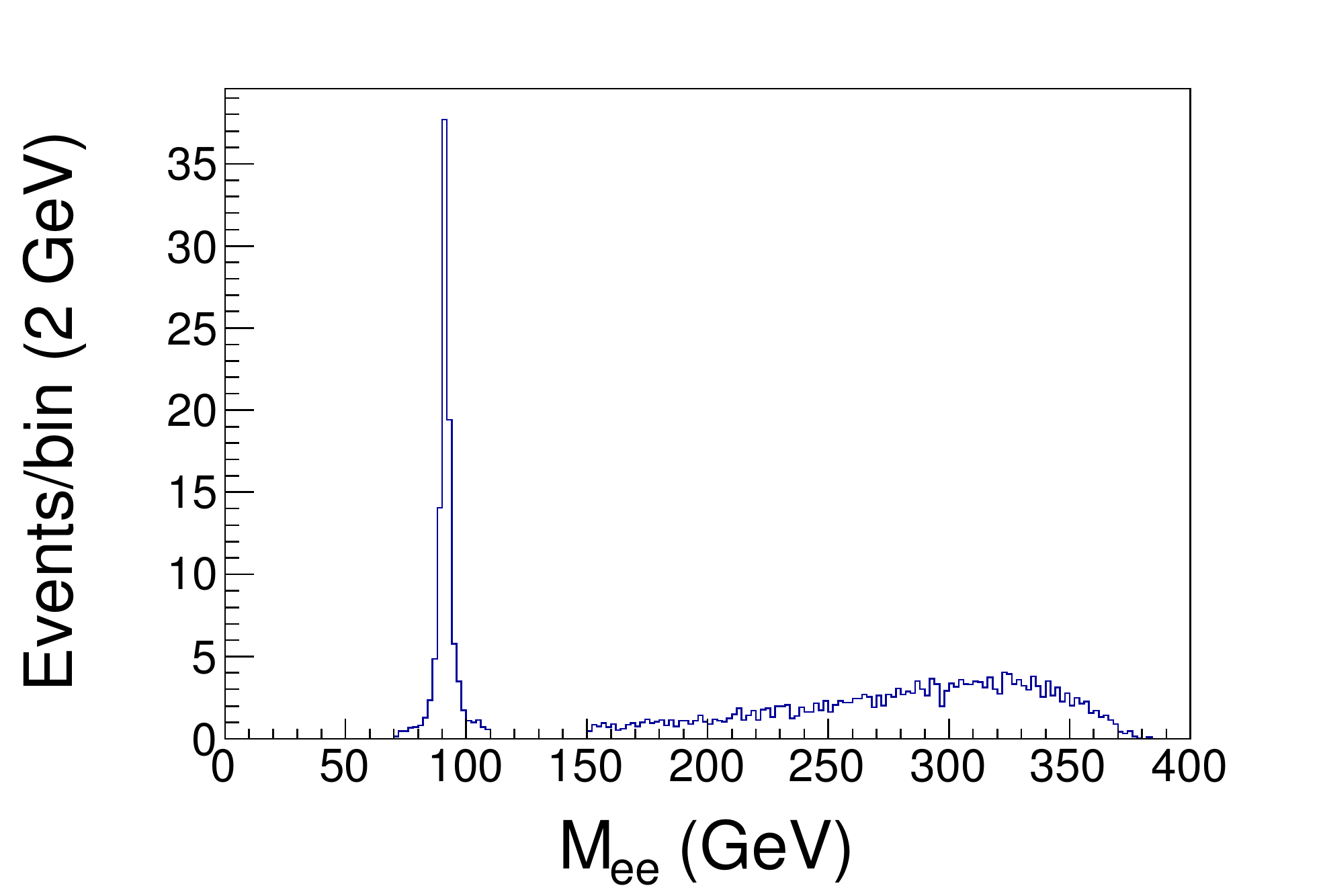}}}
\caption[]{Kinematic distributions for the signal $e^+e^- \rightarrow e^+e^-h \rightarrow e^+e^-W^+W^- \rightarrow e^+e^-4j$, after the hadronic decay selection criteria are applied for
$500$ fb$^{-1}$ at $500$ GeV. In the left panel: the recoil mass using outgoing electrons; in the right panel: invariant mass of the recoil electron pair.}
\label{fig:2}
\end{figure}

For our event selection, we first require that either $70 <M_{ee} < 110$ GeV, which we consider a Higgstrahlung event, or $M_{ee}>150$, which we take as a fusion event.
The distributions for the recoil mass and the invariant mass for our signal are shown in Figure \ref{fig:2}, one can clearly see the Higgstrahlung and fusion regions in the latter.

If there are more than 3 initial jets, and $M_{h}>115$ GeV, and $E_{miss} < 50$ GeV we treat the event as a fully hadronic decay. Otherwise we consider it as semi-leptonic and require that it have at least two hadronic jets and one additional electron or muon track.

\subsubsection{Fully Hadronic Reconstruction}
\begin{figure}[tb]
\centering
\includegraphics[angle=0,scale=0.4]{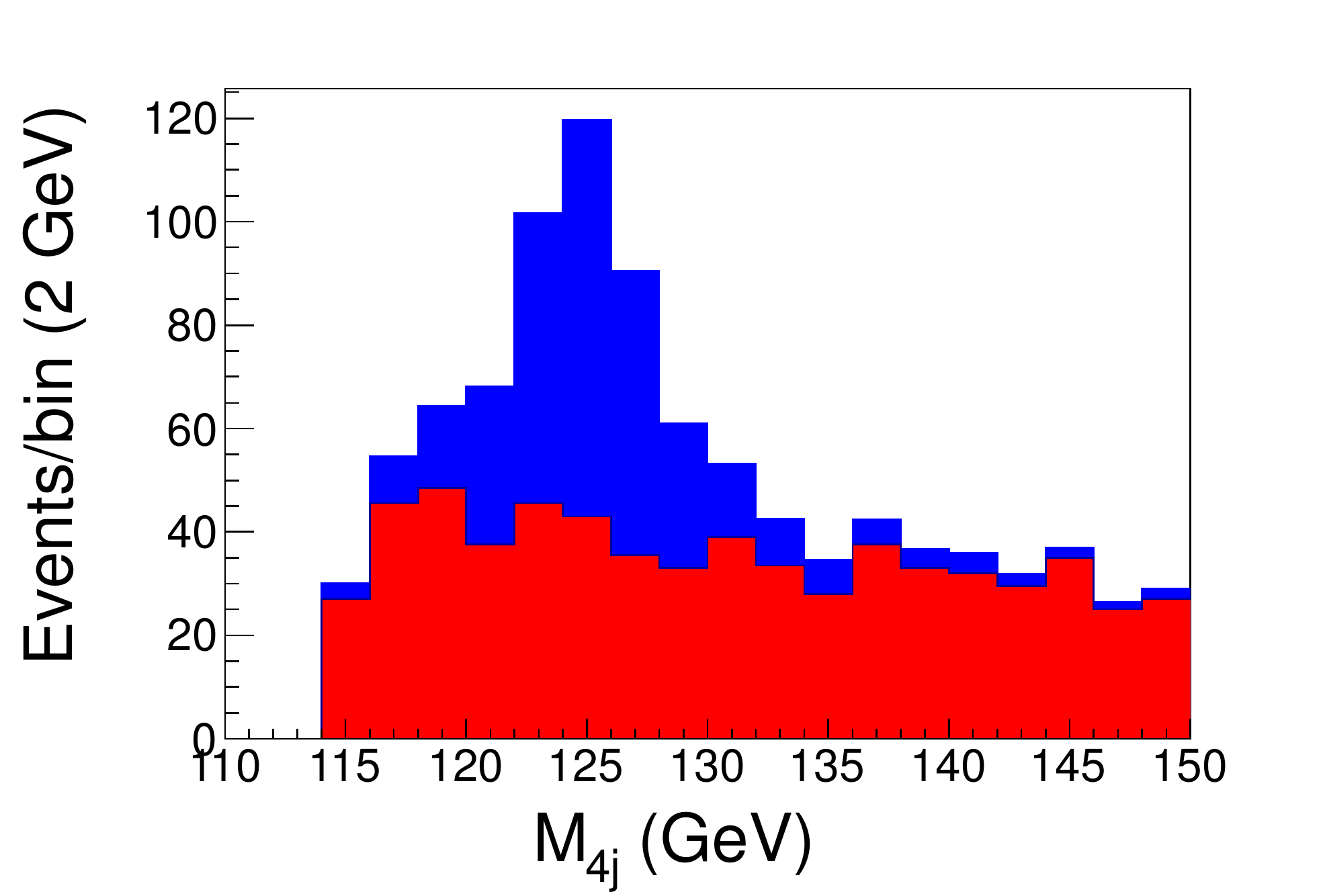}
\caption{Invariant mass of the 4 jets identified with a Higgs in the all-hadronic analysis for $500$ fb$^{-1}$ at $500$ GeV. The Higgs signal is shown in red on top of the primary background arising from
$e^+e^- \rightarrow e^+e^-q\overline{q}$ in blue.}
\label{fig:m4j}

\end{figure}

For the fully hadronic case we merge the existing jets according to the Durham algorithm until there are only four. The Durham jet definition is a sequential combination algorithm which merges the nearest sub-jets at each step according to the distance parameter
\bea
Y \equiv 2~\text{min}[E_1^2,E_2^2](1-\cos \theta_{12}).
\eea
We denote
by $Y_{45}$ the distance parameter at which the fifth jet is merged into the fourth and similarly for $Y_{34}$. We then take the pair of jets which has an invariant mass closest to the physical $W$ mass and treat this as the on-shell $W$. The remaining pair are then regarded as the off-shell $W$. The sum of the two $W$ momenta is identified with the Higgs, with corresponding
mass $M_h$. Figure~\ref{fig:m4j} shows a simulation of the signal $M_h$ along with the dominant background.

The six outgoing particles of the signal provide a number of angular variables which can be useful in discriminating against background. The decay of a Higgs through a pair of $W$s has been analyzed in detail in Ref.~\cite{Dobrescu:2009zf} and we adopt the angular variables described therein. We first boost to the rest frame of the Higgs. Then for each $W$ we boost to the rest
frame of that $W$ and compute the angle between one of the jets from its decay and the boost direction of the $W$ with respect to the Higgs rest frame. We choose whichever jet gives an angle less than $\frac{\pi}{2}$ and call these angles $\theta_{j1},\theta_{j2}$ for
the two $W$s. A third angle, $\phi_j$, describes the angle between the planes of decay of the $W$s in the Higgs rest frame.

We adopt a similar treatment for the incoming and outgoing electrons. We again begin in the Higgs rest frame and define $\theta_{l1},\theta_{l2}$ as the angles between the incoming/outgoing electron associated with each $Z$, and the boost direction of the $Z$ with respect to the Higgs, in the $Z$ rest frame. Then $\phi_l$ is the angle between the incoming and outgoing lepton pair ``decay'' planes\footnote{There is a seventh physical angle describing the orientation of the Higgs decay relative to the $Z$s involved in producing it, but this does not seem to show any useful structure in the backgrounds or signal. This is expected for the signal since the scalar Higgs cannot transmit angular correlation information.}.

With these reconstructions and definitions, we impose the following cuts: The recoil mass must be between $110$ and $250$ GeV. The 4-jet reconstructed Higgs mass, $M_{had}$ must be less than $150$ GeV. We choose these cuts because the recoil mass provides a sharper resolution at the low mass edge while the jet reconstruction is better for the high mass cut-off. The off-shell reconstructed $W$ must have an invariant mass less than $70$ GeV and its momentum in the rest frame of the Higgs ( $|P_{W}^{rest}|$) must be less than $45$ GeV. The on-shell $W$ should have an invariant mass between $55$ and $100$ GeV. For further discrimination against the backgrounds we rely on a likelihood function $L$. This function differs in the Higgstrahlung ($L_h$) and $ZZ$ fusion ($L_f$) analysis regions.
$L_h$ and $L_f$ take as inputs $\theta_{j1},\theta_{j2},\theta_{l1},\theta_{l2}, \phi_{j},\phi_{l}, |P_{W}^{rest}|, Y_{34}, N_{hj}, M_{W}^{off}$, and $M_{h}$. $L_f$ also uses $M_{ee}$. These functions are defined as the logarithm of the ratio of background to signal probability distributions in
the input variables. We have only used simple functions, such as Gaussians and exponentials, to approximately fit these distributions and have not tried to include complicated correlations, so a more detailed analysis might improve their efficacy.

\subsubsection{Semi-leptonic Reconstruction}
For the semi-leptonic decays, we proceed in analogous fashion. We require at least one additional electron or muon candidate and take the highest energy track among those found as our decay-product lepton. (Disregarding the two which are selected as recoiling electrons.) As before, we subtract this track from a hadronic jet if necessary. We identify the missing energy and momentum with
the neutrino. We then merge the hadronic jets down to two, discarding the event if there is initially only one. The sum of the hadronic jets is considered to be one $W$ while the other is the sum of the charged lepton and the neutrino. The Higgs is then
the sum of the two $W$s and $W^{on}$ is whichever has an invariant mass nearer the physical $W$ mass. Angles are defined as in the all-hadronic case except that, for the leptonically decaying $W$, $\theta_{j2}$ and $\phi_j$ are defined by the charged lepton instead of the nearer jet to the boost direction in the $W$ rest frame. We do not attempt to reconstruct the tau decays. The Higgstrahlung and fusion regions are defined as before and we apply the following cuts: The recoil mass must be greater than $115$ GeV. The candidate Higgs mass, constructed from two jets plus a charged lepton plus missing energy, must be between $100$ and $150$ GeV. The off-shell $W$ should have an invariant mass between $10$ and $60$ GeV. Further cuts are imposed by likelihood functions, which depend on $\theta_{j1},\theta_{j2},\theta_{l1},\theta_{l2}, \phi_{j},\phi_{l}, M_{W}^{off}$, and $M_{h}$ and, for the fusion case, $M_{ee}$. Our cuts are summarized in Table \ref{tab:cuts}.

\begin{table}[tb]
\centering
\begin{tabular}{|c|c|c|c|c|}
\hline
Variable & \multicolumn{2}{|c|}{Hadronic} & \multicolumn{2}{|c|}{Semi-Leptonic} \\ \hline
& Higgstrahlung & Fusion & Higgstrahlung & Fusion\\ \hline \hline
$M_{rec}$ &\multicolumn{2}{|c|}{$>110,<250$} & \multicolumn{2}{|c|}{$>115$} \\ \hline
$M_h$ &\multicolumn{2}{|c|}{$<150$} & \multicolumn{2}{|c|}{$>100,<150$}\\ \hline
$M_{W}^{off}$ & \multicolumn{2}{|c|}{$ < 70$} & \multicolumn{2}{|c|}{$>10, < 60$} \\ \hline
$|P_{W}^{rest}|$ & \multicolumn{2}{|c|}{$ < 45$} & \multicolumn{2}{|c|}{} \\ \hline
$M_{W}^{on}$ &\multicolumn{2}{|c|}{$>55, < 100$} & \multicolumn{2}{|c|}{}\\ \hline
$L$ & $L_b$ & $L_f$ & $L_b$ & $L_f$\\ \hline
\end{tabular}
\caption[]{Cuts applied in hadronic and semi-leptonic analysis for $e^+e^- \rightarrow e^+e^- h \rightarrow e^+e^-W^+W^-$.}
\label{tab:cuts}
\end{table}

\begin{table}[tb]
\centering
\begin{tabular}{|c|c|c|c|c|}
\hline
Process & \multicolumn{4}{|c|}{Expected Events} \\ \hline
& \multicolumn{2}{|c|}{All Hadronic} & \multicolumn{2}{|c|}{Semi-leptonic} \\ \hline
& Higgstrahlung & $ZZ$ Fusion & Higgstrahlung & $ZZ$ Fusion \\ \hline
$eeh \rightarrow ee4j$ & $85$ & $183$ & $0$ & $0$ \\ \hline
$eeh \rightarrow eeqql\nu$ & $1$ & $1$ & $52$ &$111$ \\ \hline \hline
$ee \rightarrow eeqq$ & $65$ & $100$  & $39$ & $59$\\ \hline
$ee \rightarrow ee4q/eeqqgg$ & $38$ & $85$ & $2$ & $1$\\ \hline
$ee \rightarrow eeqql\nu$ & $1$ & $9$ & $8$ & $41$ \\ \hline
$ee \rightarrow qql\nu$ & $<18$ & $<18$ & $<18$ & $<18$ \\ \hline \hline
Total Background &$104(+18)$ & $194(+18)$ & $49(+18)$ & $101(+18)$ \\ \hline
$\delta \sigma$ & $16\%$ & $11\%$ & $19\%$ & $13\%$ \\ \hline
\end{tabular}
\caption[]{Expected number of events from $h \to WW$ signal and backgrounds with $500$ fb$^{-1}$ at $500$ GeV ILC.}
\label{tab:wwevents}
\end{table}

\subsubsection{Sensitivity}

To estimate our sensitivity we include a number of backgrounds which are expected to contribute significantly after cuts. We model the processes $e^+e^- \rightarrow e^+e^-q\overline{q}$, $e^+e^- \rightarrow q\overline{q}l\nu$,$e^+e^- \rightarrow e^+e^-jjjj$, and $e^+e^- \rightarrow e^+e^-q\overline{q}l\nu$. Among these $e^+e^- \rightarrow e^+e^-q\overline{q}$ is the most significant in both the fully hadronic and semi-leptonic channels. Not surprisingly, $e^+e^- \rightarrow e^+e^-4j$ also contributes significantly to the fully hadronic analysis background and $e^+e^- \rightarrow e^+e^-q\overline{q}l\nu$ to the
semi-leptonic background. We provide the expected number of events from various sources which pass our cuts in Table \ref{tab:wwevents} below, assuming $500$ fb$^{-1}$ of integrated luminosity.

The last row before the total background sum sets an upper limit on any background contributions from $e^+e^- \rightarrow qql\nu$. No events in our generated sample for this process pass the cuts, but due to the large initial cross section we are not sensitive to a number of observed events smaller than $\sim 18$.

A further consideration is the effect of Higgs decays to $b$-quark pairs, which are expected to present a large branching fraction for a SM-like scalar. These decays can potentially pass our cuts and contribute to the excess over non-Higgs backgrounds but they could limit our ability to measure the pure $WW$ signal. However, this concern can be largely addressed
with b-tagging techniques. We do not include an explicit $b$-tagging simulation in our analysis, however, reasonable estimates show that the net effect of $b$-quark decays and $b$-tagging is small and we will proceed based on the assumption that this background can be neglected. See Appendix~\ref{sec:bkg} for a more detailed discussion of these effects.

Based on the numbers above we estimate the sensitivity to the combined all-hadronic and semi-leptonic signals using
 \bea
\delta \sigma_{eeh \to eeWW} = \frac{\sqrt{N_S + N_B}}{N_S}
\label{eq:width5}
\eea
where $N_S$ and $N_B$ are the expected number of signal and background events respectively. This gives an uncertainty on the signal cross section $\delta \sigma_{eeh \to eeWW} = 6.8 \%$. However, we can additionally make use of Higgstrahlung signal events where the $Z$ decays to muons. This should give us essentially the same number of signal events as the Higgstrahlung to electrons channel. If we very conservatively assume that the background is also the same, we can bring the error down to
$\delta \sigma_{eeh \to eeWW} = 6.0 \%$. Assuming no new background would bring our error down to $5.6 \%$. For the results below we will use $\delta \sigma_{ZW} = 6 \%$ for this channel.


\section{Achievable Accuracies at the ILC}
\label{sec:accur}

\subsection{Model-Independent Fitting}
\label{sec:mi}
The expected statistical uncertainty on various other cross sections at the ILC have been calculated by several authors~\cite{Baer:2013cma,Asner:2013psa}. We will make use of numbers presented for the Snowmass Community Study for cross sections other than those we have calculated ourselves. Table~\ref{tab:dbd} lists the ILC uncertainties we use in our analysis below. Table~\ref{tab:LHC} lists the most relevant LHC uncertainties in our study. These include assumptions about future theoretical and systematic errors. ATLAS and CMS use different extrapolation assumptions to obtain their high luminosity precisions. We have combined them in a conservative way to estimate the effect of both experiments, for details of the combination in Appendix.~\ref{sec:chidef}.

\begin{table}[tb]
\centering
\begin{tabular}{|c|c|c|c|c|c|}
\hline
$\sqrt s$ and $\mathcal{L}$ &  \multicolumn{2}{|c|}{ILC250} & \multicolumn{2}{|c|}{ILC500} & \multicolumn{1}{|c|}{ILC1000}\\
$(P_{e^-}, ~P_{e^+})$ & \multicolumn{2}{|c|}{$(-0.8,~+0.3)$} & \multicolumn{2}{|c|}{$(-0.8,~+0.3)$} & \multicolumn{1}{|c|}{$(-0.8,~+0.2)$} \\ \hline
Decay$\backslash$ Production& $Zh$& $\nu\nu h$ & $Zh/eeh$& $\nu\nu h$ & \multicolumn{1}{|c|}{ $\nu\nu h$} \\ \hline
$inclusive$ (\%)& 2.6 & & 3.0 & &\\ \hline
 $b\overline{b}$ (\%)& $1.2$ &  $11$ & $1.8$ & $0.66$ & $0.5$\\ \hline
$c\overline{c}$ (\%)& $8.3$ &  & $13$ & $6.2$ & 3.1\\ \hline
$\tau^+\tau^-$ (\%)& 4.2 & & 5.4 & 9.0 & 2.3 \\ \hline
$gg$ (\%)& $7$ & & $11$ & $4.1$ & 1.6 \\ \hline
$WW$ (\%)& $6.4$ & & {\bf 6} & $2.4$ & 3.1 \\ \hline
$ZZ$ (\%) & $19$ & & $25$ & $8.2$ & 4.1 \\ \hline
$\gamma \gamma$ & $29-38$ & & $29-38$ & $20-26$ & 8.5 \\ \hline \hline
$\br_{inv} (\%)$ & {\bf 0.5-0.7} & & & & \\ \hline
\end{tabular}
\caption[]{Estimated relative errors for various cross sections at the ILC. The first and second rows indicate the production energy and mechanism respectively
for the Higgs while the first column shows the decay modes. Numbers are taken from the ILC Snowmass Whitepaper \cite{Asner:2013psa} except for simulation studies presented in
this paper. The Br$_{inv}$ is absolute error.
}
\label{tab:dbd}
\end{table}
\begin{table}[tb]
\centering
\begin{tabular}{|c|c|c|c|c|c|c|}
  \hline
             &          & $\gamma\gamma$ (\%)& $WW^*$ (\%)& $ZZ^*$ (\%)& $b\bar b$ (\%)& $\tau^+\tau^-$ (\%)\\ \hline
     LHC          & ATLAS    & 9-14         & 8-13 & 6-12 & N/A       & 16-22 \\ \cline{2-7}
  $300~\fbi$ & CMS      & 6-12         & 6-11 & 7-11 & 11-14   & 8-14  \\ \cline{2-7}
           & Combined & {\bf 12}           & {\bf 10}   & {\bf 10}   & {\bf 14}      & {\bf 14} \\ \hline
   HL\_LHC            & ATLAS    & 4-10         & 5-9  & 4-10 & N/A       & 12-19 \\ \cline{2-7}
 $3000~\fbi$ & CMS      & 4-8          & 4-7  & 4-7  & 5-7     & 5-8 \\ \cline{2-7}
           & Combined & {\bf 7}            & {\bf 7}    & {\bf 6}    & {\bf 7}       & {\bf 8} \\
  \hline
\end{tabular}
\caption[]{Relative uncertainties of relevant quantities from projections of ATLAS and CMS experiments for LHC 14 TeV with $300~\fbi $ and $3000~\fbi $ (HL\_LHC) integrated luminosity from Snowmass studies~\cite{Dawson:2013bba}. See the text and Appendix~\ref{sec:chidef} for combination details.
}
\label{tab:LHC}
\end{table}

\begin{figure}[tb]
\centering
\includegraphics[width=300pt]{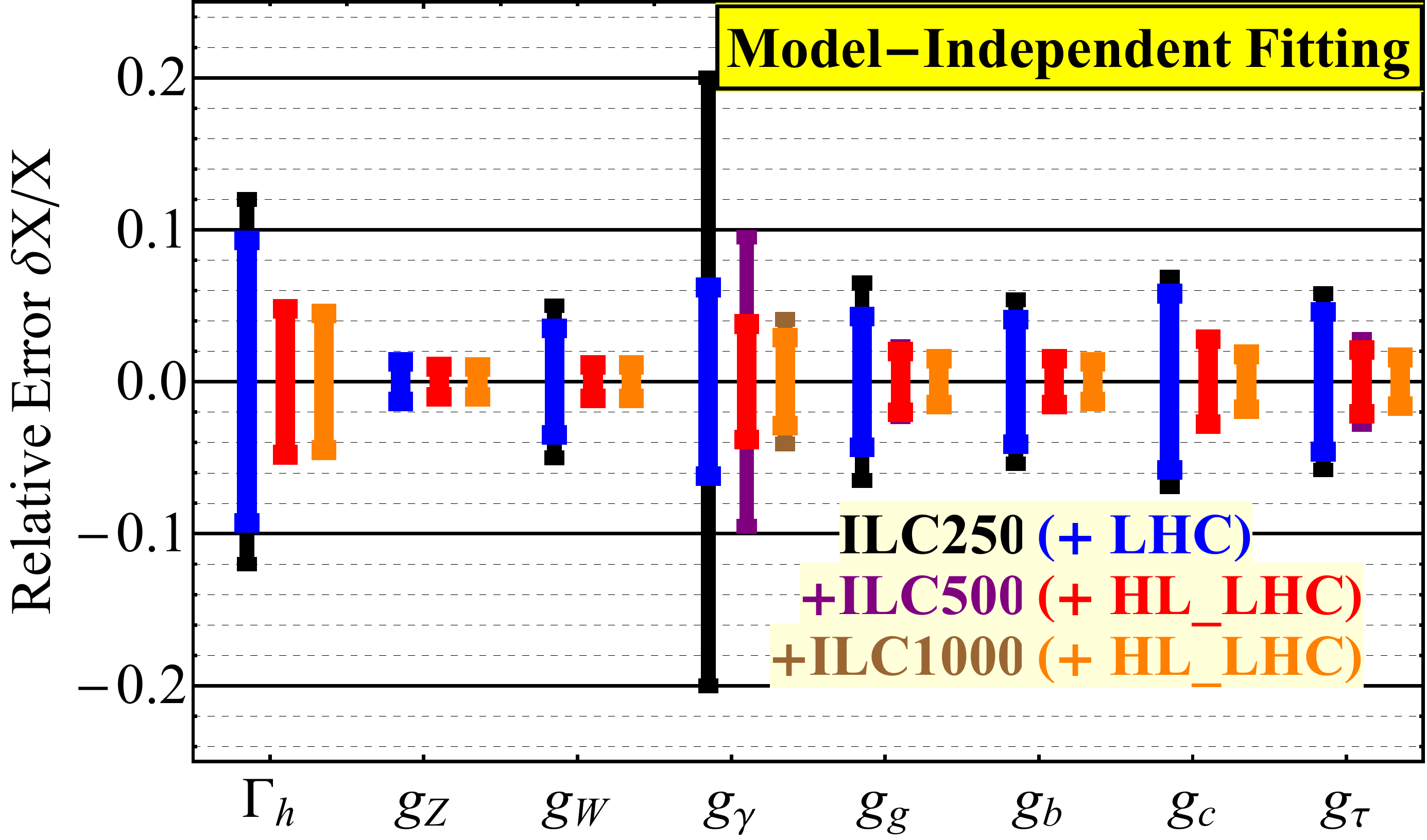}
\caption{Model-independent uncertainties of the Higgs boson couplings from ILC $250~\gev$ with (blue) and without (black) LHC data with $250~\fbi$ of integrated luminosity; at $500~\gev$ (red/purple) with $500~\fbi$; and at $1~\tev$ (orange/brown) with $1000~\fbi$. Results at $250~\gev$ are combined with LHC 14 TeV using $300~\fbi$ projections while those at $500~\gev$ and $1~\tev$ use $3000~\fbi$ projections.  For details see Table.~\ref{tab:mi} and the text.
}
\label{fig:mi}
\end{figure}

Our procedure for this model-independent fit is described in Appendix~\ref{sec:chidef}, especially in Eq.~(\ref{eq:chi1}). This fit also determines the relative error on the various coupling constants $g_A$. The results are given below in Table \ref{tab:mi} and Figure~\ref{fig:mi}. We present the expected errors at the $250$, $500$ and $1000$ GeV stages of running with integrated luminosities of $250~\fbi$, $500~\fbi$, and $1000~\fbi$ respectively, henceforth labeled as ILC250, ILC500 and ILC1000 scenarios. For each scenario we include projected sensitivities with and without the addition of information from the LHC. For the invisible decays we present the case for a $10\%$ invisible branching fraction and for $1\%$. As mentioned in Sec.~\ref{sec:hinv}, the addition of the invisible search does not constrain the other Higgs couplings or total width in a model-independent fit.

\begin{table}[tb]
  \centering
    \begin{tabular}{|c|cc|cc|cc|}
    \hline
    \multirow{1}[0]{*}{Relative Error} & \multicolumn{2}{|c|}{\multirow{2}[0]{*}{ILC250}} & \multicolumn{2}{|c|}{\multirow{2}[0]{*}{+ILC500}} & \multicolumn{2}{|c|}{\multirow{2}[0]{*}{+ILC1000}}  \\
      \%    &  & &  & & & \\ \hline
    $\Gamma$ & 12 & (9.3) & 4.8 & (4.8)  & 4.5 & (4.5) \\ \hline
    $g_Z$  & 1.3 & (1.3) & 0.99 & (0.99) & 0.98 & (0.98)\\ \hline
    $g_W$  & 5.0 & (3.5) & 1.1 & (1.1) & 1.1 & (1.1) \\ \hline
    $g_\gamma$ & 20 & (6.2) & 9.5 & (3.8) & 4.1 & (2.9) \\ \hline
    $g_g$  & 6.5 & (4.3) & 2.3 & (2.0)  & 1.5 & (1.5)\\ \hline
    $g_b$  & 5.4 & (4.1) & 1.5 & (1.5) & 1.3 & (1.3) \\ \hline
    $g_c$  & 6.9 & (5.8) & 2.8 & (2.8)  & 1.8 & (1.8) \\ \hline
    $g_\tau$ & 5.8 & (4.6) & 2.8 & (2.1)  & 1.6 & (1.6) \\ \hline\hline
    $\br_{inv} (\%)$ & \multicolumn{6}{|c|}{0.6~(0.5)}\\ \hline
    \end{tabular}%
\caption[]{Model-independent precisions ($1\sigma$) of the width and couplings constants expected for a SM-like Higgs at
three stages of ILC run. All results assume completion of previous stage of ILC runs. Results in combination with LHC projections are in parenthesis.
We combine $300$ fb$^{-1}$ estimates for the LHC with the $250$ GeV ILC run, and $3000$ fb$^{-1}$ with the $500$ GeV run and $1~\tev$ run.
The absolute value of uncertainty on $\br_{inv}$ is given for an input $\br_{inv}$ of 10\% (1\%).
}
\label{tab:mi}
\end{table}

Using the $\sigma_{ZW}$ determination above we can compute the Higgs width in terms of Eq.~(\ref{eq:width4}). $\sigma_{ZW}$ can be further constrained by measurements of the process $e^+e^- \rightarrow Zh, h \rightarrow WW$ at $250$ GeV and $e^+e^- \rightarrow \nu\nu h, h \rightarrow ZZ$ at $500$ GeV. Combining the information from these channels we estimate that $\sigma_{ZW}$ can be determined to a relative error of $\delta_{ZW} = 4.6 \%$. Using the available estimates for $\delta\sigma_{Zb}$ and $\delta\sigma_{Wb}$ one finds that the total error on $\Gamma_h$ using Eq.~(\ref{eq:path250}) is $6.8 \%$. This is nearly the same as the error achievable using Eq.~(\ref{eq:width4}).

As can be seen from Table.~\ref{tab:mi} and Figure.~\ref{fig:mi}, the results for ILC alone fitting are comparable with the model-independent fitting in Ref.~\cite{Dawson:2013bba}. Our study for the invisible Higgs decays in Sec.~\ref{sec:inv} and $ZZ$-fusion Higgs to $WW$ both improves the ILC measurements when comparing Table.~\ref{tab:dbd} with Ref.~\cite{Baer:2013cma}.
The combination with LHC measurements from Table.~\ref{tab:LHC} improves almost all couplings precisions for 250 GeV ILC. The improvement for total width in this case receives contributions from several channels at the LHC as described in Eq.~(\ref{eq:path3}).

Especially for $g_\gamma$ and $g_g$, which are statistically limited at the ILC, the LHC can provide sizable gains. With ILC 500 data many of these benefits become marginal due to increased sensitivity from ILC channels alone. Still, the improvement in the $g_\gamma$ remains significant. Including an ILC run at $1~\tev$ leads to further gains in the fermion and photon channels and essentially obviates the effect of LHC information.

As noted before, the largest contribution to the total width error derives from the uncertainty on the inclusive cross section. Further studies on $ZZ$-fusion inclusive measurement at the ILC for $500$ GeV and $1$ TeV should be valuable. Approximately half of the expected sensitivity ($\sim 3 \%$) on the inclusive measurement at $500$ GeV comes from Higgstrahlung events with $Z$ decaying to electrons or muons. If the fusion cross-section, which is roughly twice as many events, can be utilized efficiently the overall precision at $500$ GeV might be pushed down to nearly the $2 \%$ level, which in combination with the $250$ GeV measurement could determine the inclusive cross-section at less than $2 \%$.


\subsection{Model-Dependent Constraints}
\label{sec:md}

Thus far we have proceeded with a strictly model-independent method for fixing the couplings, treating $\Gamma_h$ as an independent parameter. Effectively, this means we allow for arbitrarily large Higgs decays into ``buried'' channels which are not
constrained at the ILC. One may improve on the results by adding the reasonable assumption that any buried channels in the clean environment of the ILC are negligibly small compared to the total width, i.e., if we assume that the total width is the sum over partial widths arising from the coupling constants fitted above. This would be true for a SM-like Higgs. By including the search for invisible decays we can make this assumption considerably more robust.

\begin{table}[tb]
\centering
\begin{tabular}{|c|c|c|c|c|c|c|c|}
\hline
& SM Theo. & \multicolumn{2}{|c|}{ILC250} & \multicolumn{2}{|c|}{+ILC500} & \multicolumn{2}{|c|}{+ILC1000} \\ \cline{3-8}
& Error on & MDA & MDB  & MDA & MDB & MDA & MDB\\
& $\br(\%)$ & ($\pm$\%) & ($\pm$\%) &($\pm$\%) & ($\pm$\%) &($\pm$\%) & ($\pm$\%) \\ \hline
${\Gamma_h}$  & $+3.9,-3.8$ & 1.5 & 7.8 & 0.84 & 4.4 & 0.67 & 4.2 \\ \hline
$g_Z$ & $\pm4.2$ & 0.75 &\multirow{2}[0]{*}{1.3} & 0.44 & \multirow{2}[0]{*}{0.99}  & 0.41 & \multirow{2}[0]{*}{0.98}\\ \cline{1-3} \cline{5-5} \cline {7-7}
$g_W$ & $\pm4.1$ & 2.8 & & 0.38 &   & 0.25 & \\ \hline
$g_\gamma$ & $\pm4.9$ & 6.0 & 6.1 & 3.6  & 3.8 & 2.7 & 2.9\\ \hline
$g_g$  & $\pm10.1$ &2.7 & 3.9 & 1.5 & 1.9 & 0.89 & 1.4 \\ \hline
$g_b$ & $\pm3.4$  &1.4 & 3.3 & 0.75 & 1.4 & 0.59 & 1.3 \\ \hline
$g_c$ & $\pm12.2$ &4.3 & 5.2 & 2.5 & 2.7 & 1.4 &1.8 \\ \hline
$g_\tau$ & $\pm5.6$  &2.3 & 3.7 & 1.6 & 2.0  & 1.1 & 1.5 \\ \hline
${\rm Br}_{inv}$ & --- & $<0.52$ & 0.60 & $<0.52 $ & 0.57 & $<0.52$& 0.57\\ \hline
\end{tabular}
\caption[]{Model-dependent precisions for coupling constants achievable at the ILC, combined with LHC (HL\_LHC) under two different assumptions for ILC250 (ILC500 and ILC1000). MDA assumes no invisible decays above background. MDB assumes $g^2_{hWW}/g^2_{hZZ}={\rm cos}^2\theta_{\rm w}$. Note that in MDA $\Gamma_h$ is no longer a free parameter, and in MDB $g_W$ and $g_Z$ are essentially the same parameter, $g_V$.  SM theoretical uncertainties are shown in the second column, from Ref.~\cite{Dittmaier:2011ti,Dittmaier:2012vm}. Uncertainty on $\br_{inv}$ is absolute value and input of $\br_{inv}$ is set at $10\%$.
}
\label{tab:md}
\end{table}

\begin{figure}[tb]
\centering
\subfigure{
\includegraphics[width=300pt]{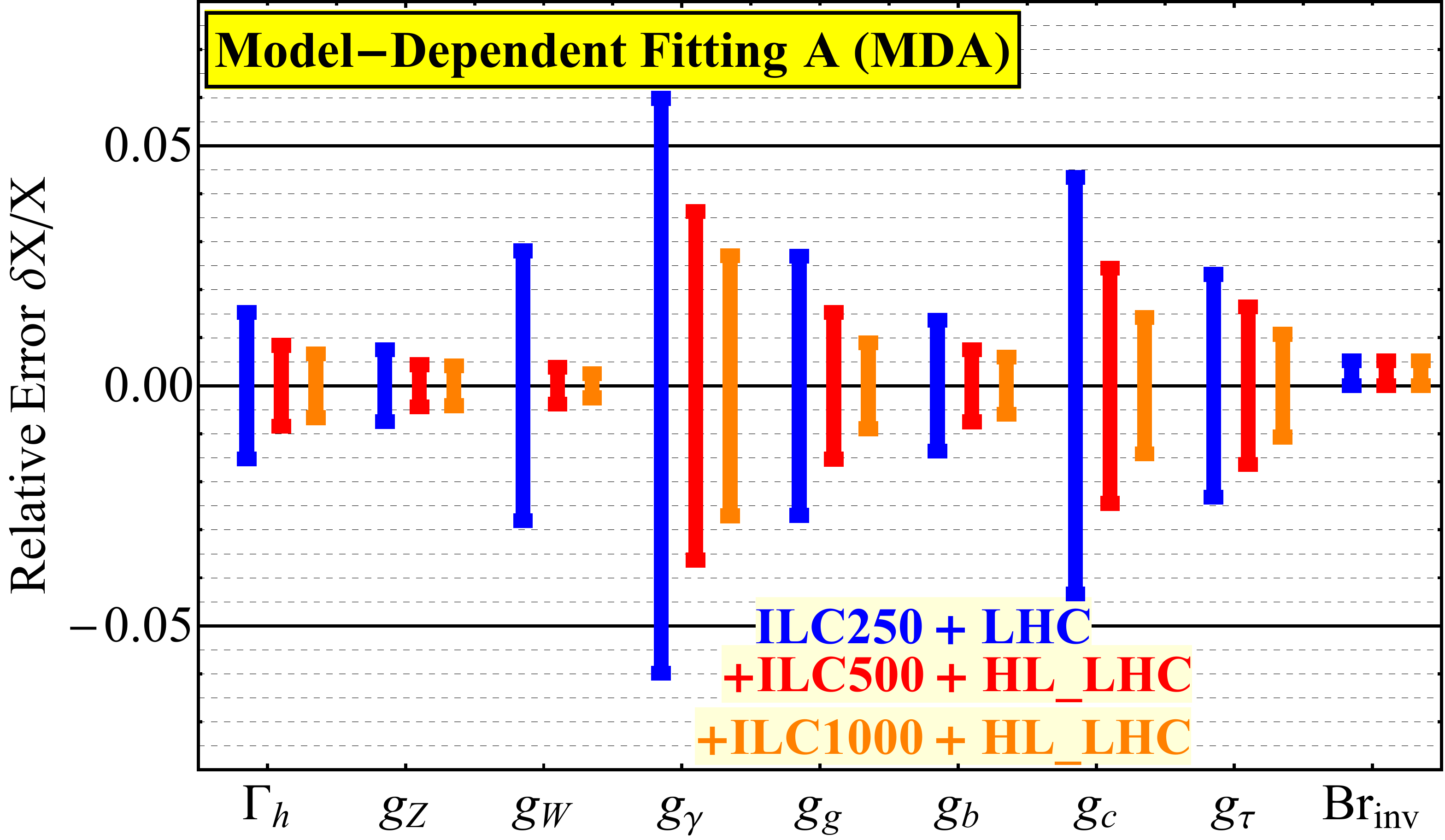}}
\subfigure{
\includegraphics[width=300pt]{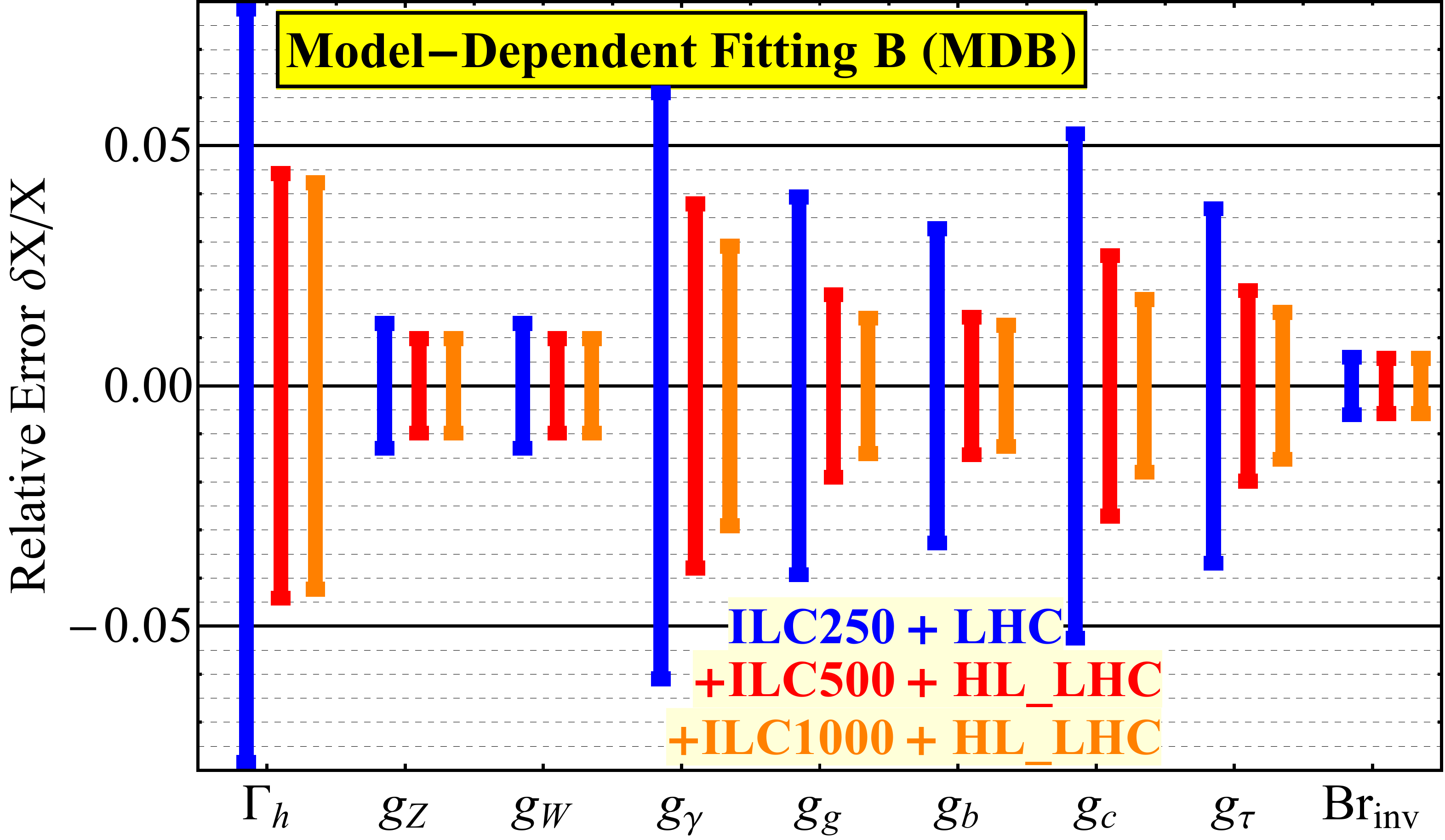}}
\caption[]{Model-dependent uncertainties of the Higgs boson couplings from ILC250 (blue), ILC500 (red) and ILC1000 (orange) runs under assumption MDA (first panel) and MDB (second panel). Estimates include projected information from the LHC. For details, see caption of Table.~\ref{tab:md} and the text. Uncertainty on $\br_{inv}$ is absolute value and input of $\br_{inv}$ is set at $10\%$.

}
\label{fig:md}
\end{figure}

In Table~\ref{tab:md} and Figure~\ref{fig:md} we present expected sensitivities with the assumption that, including the invisible channel, all significant decay modes are observable. We include LHC information in Table~\ref{tab:LHC} for all columns. Two cases are shown for each energy:

\begin{itemize}
\item[1.]
MDA (Model-Dependent Case A) shows the case where no excess is observed and only an upper limit on the invisible decay cross section can be set. Since in this case the invisible signal is consistent with zero, we only show the upper limit on the branching fraction.
\item[2.]
MDB (Model-Dependent Case B) shows the case assuming a tree-level custodial symmetry relation $g^2_{hWW}/g^2_{hZZ} = {\rm cos}^2\theta_{\rm w}$.\footnote{This condition holds for Higgs singlet and doublet models. However, this condition does not necessarily hold for triplet models with custodial symmetry~\cite{Georgi:1985nv,Logan:2010en}.}
\end{itemize}

MDA fitting results show strong improvement in the coupling precisions compared to the model-independent extraction as well as a large reduction on the total width error, with gains at both energy scales. We note here for the MDA fit the total width is no longer a fitting parameter, but rather a derived quantity as shown in Eq.~(\ref{eq:width}). The rarest channels show the least improvement comparing to the Model-Independent fit, which is to be expected since they are least sensitive to the overall width. This assumption, that  ``buried'' channels are a negligible contribution is valid in many models. Still, searches for exotic Higgs decays are well-motivated~\cite{exotic} and should start with a concerted effort
given the clean environment at a lepton collider where many can be explicitly ``unburied".
MDB fitting also shows improved precisions for the $250$ GeV run. Most obviously, the $g_W$ precision is set to the same level as the $g_Z$, and this in turn better constrains other couplings to quarks and leptons. The improvement on the total width is not as dramatic as in MDA since the large branching fraction to $bb$ and other modes still gives a significant contribution. With the addition of ILC $500$ information MDB has little effect on the expected precisions. That is, the ILC model-independent measurements of $g_W$ and $g_Z$ are already comparable and small so the assumption that they are equal does not affect the fit much. As in the model-independent approach, ILC1000 can improve the sensitivity for fermionic, photonic and gluon couplings in MDA or MDB.
We list in the second column the theoretical uncertainties on the total width $\Gamma_h$ and branching fractions of a $126$ GeV SM Higgs from Ref.~\cite{Dittmaier:2011ti,Dittmaier:2012vm}. Roughly twice the uncertainty on coupling constants enters into the Brs. One can see that for both MDA and MDB the statistical precisions on couplings are comparable to theoretical uncertainties. In principle the theory errors are reducible but effort will be needed to make maximum use of the potential at a Higgs factory.

\section{Summary and Conclusions}

In this paper we have outlined a systematic approach to the determination of the Higgs total width
and measurable coupling parameters in a model-independent manner at the ILC in Sec.~\ref{sec:approach}, and illustrated the complementarity for operating the ILC at $250~\gev$, $500~\gev$ and $1~\tev$.
We have performed detailed simulations for two channels which can play an important role in determining the total width with high precision in model-independent and model-dependent scenarios.
We first included the invisible decay channels at  $250~\gev$, and carried out new analyses of the $e^+e^- \rightarrow e^+e^- h$ with $h \rightarrow \ww$ at $500~\gev$.
In Sec.~\ref{sec:accur}, we estimated the achievable accuracies on coupling constants and total width at the ILC. We also emphasized the benefits and importance of combining measured cross sections from the LHC and the ILC, which results in improved precision, especially for the total width at $250~\gev$ ILC and precisions for $g_\gamma$ and $g_g$ for $250~\gev$ as well as $500~\gev$ ILC. With the statistics assumed for a $1$ TeV ILC run, LHC information leads to only small improvements.

Our specific findings can be summarized as follows.
\begin{itemize}
\item[1.]
At $250$ GeV the key measurement of the inclusive Higgstrahlung cross section, which enters all the partial width determinations in this approach, can be made with high precision as discussed in Sec.~\ref{sec:250}. However, the precision on the total width is limited by the error on other exclusive cross sections, such as $\ee \rightarrow \ee h \rightarrow \ee \zz$ and $\ee \rightarrow \nu\nu h \rightarrow \nu\nu \ww$. This is where the additional information for the coupling ratios from the LHC provides important enhancement of the achievable precision as discussed in Sec.~\ref{sec:mi}. As shown in Table~\ref{tab:mi}, any couplings can be measured in a fully model-independent way at this energy to the $(3-5)\%$ level. Under the minimal assumption that the searches for visible and invisible modes comprise all significant decay channels or custodial symmetry, these couplings can be measured at the $(1-3)\%$ percent level.

\item[2.]
At $500$ GeV the total width can be largely determined by measuring a few channels due to the high precision expected for Higgs decays to $b\overline{b}$ produced via Higgstrahlung and $WW$ fusion as discussed in Sec.~\ref{sec:500}. We have shown that the exclusive cross section for ZZ fusion process with subsequent decay of the Higgs boson to $WW^*$ ($\sigma_{ZW}$) can also be determined with good precision and used in place of the $\sigma_{WW}$ measurement to achieve nearly the same precision on the total width in a way which is less sensitive to additional errors in the $b\overline{b}$ decay channels as shown in Sec.~\ref{sec:500ww}.
At this energy, assuming the $250$ GeV run has been completed, one can make model-independent determinations of the coupling constants at the $(1-3)\%$ level as shown in Table.~\ref{tab:mi}. Adding the assumption that all significant modes have been seen can reduce these errors to the sub-percent level for some couplings as shown in Table.~\ref{tab:md}. At this point any further improvement of the Higgs total width is limited by the uncertainty in the inclusive cross section for $Zh$ production. Improving this key measurement would require either a longer run at $250$ GeV or detailed study at $500$ GeV. At this stage efforts to reduce theoretical uncertainties are needed to consistently interpret the experimental results in terms of theoretical parameters.

\item[3.]
 A $1$ TeV ILC run with high luminosity can improve the fermion, photon and gluon coupling measurements by $\sim 25-50\%$ except for $g_b$.
\item[4.]
Good precision for Higgs to invisible, $0.5-0.7\%$, can be reached at the $Zh$ threshold at $250~\gev$, as shown in Sec.~\ref{sec:inv}.

\item[5.]
At a higher ILC energy above 250 GeV, the fusion channels will become more important.
In particular, the inclusive $ZZ$ fusion process at higher energy could provide further improvement for the  model-independent coupling precision, and should be carefully studied with respect to the various sources of backgrounds.
\end{itemize}


\begin{acknowledgments}
We would like to thank K. Fujii, J. List, J. Reuter, T. Tanabe and J. Tian for advice on simulations. This work is supported in part by the U.S.~Department of Energy under grant No. DE-FG02-95ER40896, in part by the PITT PACC. Z.L.~is supported in part  by the LHC Theory Initiative from the US National Science Foundation under grant NSF-PHY-0969510, and in part  by an Andrew Mellon Predoctoral Fellowship from Dietrich School of Art and Science, University of Pittsburgh.

\end{acknowledgments}

\vskip 0.5cm
\noindent
\textit{\underline{\bf Note added:}}\\

After the submission of this manuscript to the arxiv, a new paper on the same subject by M.~Peskin~\cite{Peskin:2013xra} appeared. While both results are consistent with each other in general, his more optimistic conclusions are due to the following differences in our treatment: (1) Our ``Model-Independent'' fit (Sec.~\ref{sec:mi}) took the Higgs boson total width as a free parameter, while the fit in Ref.~\cite{Peskin:2013xra} made an additional assumption that $\br({\rm undetected})< 0.9\%$. Our ``Model-Dependent fit A'', assuming $\br({\rm undetected}) \simeq 0$ (Sec.~\ref{sec:md}), led to similar results to his; (2) We made a conservative choice for the HL\_LHC projections (Table \ref{tab:LHC}), effects of which are very minimal except for $g_\gamma$; (3) We did not consider the ILC luminosity upgrade at 500 GeV from $500~\fbi$ to $1600~\fbi$ and at 1 TeV from $1000~\fbi$ to $2500~\fbi$.

\appendix

\section{Combination method and $\chi^2$ Definition}
\label{sec:chidef}
We list the LHC projections for most relevant modes in Table~\ref{tab:LHC}. All ATLAS and CMS projections are shown in ranges corresponding to different assumptions about systematic and theoretical uncertainties. For ATLAS, the ranges represent projections with and without theoretical uncertainties. For CMS, the ranges represent projections with and without reductions of systematic and theoretical uncertainties. The lower range corresponds to assumptions that systematical uncertainties will scale as $1/\sqrt{\mathcal{L}}$ and theoretical uncertainties will be halved. We estimate the theoretical and systematic uncertainties based on these ranges, for example, range $a-b$ indicates theoretical uncertainty is $\sqrt{b^2-a^2}$ for ATLAS. Similarly, for CMS projections in range $a-b$, $\sqrt{b^2-a^2}$ is approximately the theoretical uncertainty plus the systematic uncertainty added in quadrature. We take the lower of these two quantities from ATLAS and CMS as an estimated systematic plus theoretical uncertainty. We combine both experimental results from the lower range to approximate the statistical gain and add in the estimated theory plus systematics term quadratically. These conservative combined result are shown in Table~\ref{tab:LHC}. As one can see, most of them are only slightly better than conservative individual experimental projections, indicating the large contribution from systematic and theoretical uncertainties. We use these combined results as our input for LHC measurement for the fitting described below. Note that some portion of the theoretical uncertainties, including PDF and scaling effects, can be cancelled when taking the ratios of measurements from the LHC. Once these projections on ratios from experiments become available, LHC input in form of ratios will further help determine the Higgs couplings.

For the model-independent fittings, we have 9 parameters, these are $\Gamma_h$, $g_{b}$, $g_{c}$, $g_{g}$, $g_{W}$, $g_{\tau}$, $g_{Z}$, $g_{\gamma}$, and $\br_{inv}$. In the absence of actual experimental data, we take the central values of the measured cross sections to be equal to their standard model values. Let $\hat{g}_A$ be the fitted parameter normalized by its standard model value.  All of these take a central value of one.

For the model-independent fittings, the $\chi^2$ we used can be expressed in Eq.~(\ref{eq:chi1}), where sum over $\sigma$ means summing over all the independently measured cross sections. These cross sections includes LHC, ILC at $250~\gev$ and $500~\gev$. $\sigma_Z^{inc}$ is the inclusive $Zh$ associated production cross section measured at $250~\gev$ ILC. For the LHC measurements, we take all of the sensitivity projections for $\gamma\gamma$, $WW^*$ and $ZZ^*$ from glu-glu-fusion and $b\bar b$ from ZH associated production. As for $\tau^+\tau^-$, we take half the sensitivity to be from glu-glu-fusion and the other half from weak boson fusion~\cite{ATLAS-CONF-2012-160,CMS-PAS-HIG-13-004}. For ILC input, we take the conservative value as well. For example, we take $\sigma_{Z\gamma}$ to be $38\%$ where the estimation is in the range of $29\%-38\%$.

\begin{equation}
\chi^2=\sum\limits_\sigma\left(\frac {1-\hat{g}_A^2 \hat{g}_B^2/ \hat{\Gamma}_h} {\delta\sigma_{AB}}\right)^2+\left(\frac {1-\hat{g}_Z^2} {\delta\sigma_Z^{inc}}\right)^2+\left(\frac {1-\hat{g}_Z^2 \hat {\br}_{inv}} {\delta\sigma_{Zh\to Z+inv}}\right)^2
\label{eq:chi1}
\end{equation}

For model-dependent fittings, we have 8 parameters, these are $\br_{inv}$, $g_{b}$, $g_{c}$, $g_{g}$, $g_{W}$, $g_{\tau}$, $g_{Z}$, and $g_{\gamma}$. Again, all of the couplings are normalized to one. $\br_{inv}$ has a central value of zero. Notice that $\Gamma_h$ is no longer a fitting parameters here, instead it is determined by the other fitting parameters, as shown in Eq.~(\ref{eq:width}). This is a result of assuming that a sum over all Br gives the normalized total width. $\chi^2$ for the model-dependent case can be written as in Eq.~(\ref{eq:chi2}).

\beq
\centering
\hat{\Gamma}_h=(\sum\limits_{i}\br_i \hat{g}_{i}^2 +(1-\sum\limits_{i}\br_i))(1-\br_{inv})+\br_{inv}
\label{eq:width}
\eeq

\beq
\chi^2 = \sum\limits_\sigma\left(\frac {1-\hat{g}_X^2 \hat{g}_Y^2/ \hat{\Gamma}_h} {\delta\sigma_{AB}}\right)^2+\left(\frac {1-\hat{g}_Z^2} {\delta\sigma_Z^{inc}}\right)^2+ \left(\frac {\hat{g}_Z^2 \br_{inv}} {\delta \sigma_{Zh->Z+inv}}\right)^2
\label{eq:chi2}
\eeq

We note here that ILC measurements include systematic but not theoretical uncertainties. For our results shown in Tables \ref{tab:mi} and \ref{tab:md}, theoretical errors should be included for a consistent comparison with models.

\section{Analysis for $b\overline{b}$ Backgrounds}
\label{sec:bkg}
A potential complication to our signal which we do not explicitly include is the ``background'' coming from $h \rightarrow b\overline{b}$ decays. Kinematically, these events are very similar to the $WW$ signal except for the details of the 4-jet substructure.
With the cuts described above, a SM-like $h\rightarrow b\overline{b}$ process would contribute events as in the Table \ref{tab:bb}.

\begin{table}
\centering
\begin{tabular}{|c|c|c|c|c|}
\hline
& \multicolumn{2}{|c|}{All Hadronic} & \multicolumn{2}{|c|}{Semi-leptonic} \\ \hline
& Higgstrahlung & $ZZ$ Fusion & Higgstrahlung & $ZZ$ Fusion \\ \hline
$eeh\to eebb$ &$100$ & $112$ & $33$ & $42$ \\ \hline
\end{tabular}
\caption[]{Additional events expected from $h \to b\overline{b}$ with the cuts described in Sec.~\ref{sec:500ww}. No $b$-tagging has been applied for this table.}
\label{tab:bb}
\end{table}

Although adding to the excess over non-Higgs SM backgrounds, these events would degrade our ability to precisely measure the $g_W$ coupling. However, we can make use of the strong $b$-tagging capabilities expected at the ILC to reduce this problem~\cite{ILCtag}. Tagging algorithms can be characterized in terms of their $b$-acceptance efficiency, $\epsilon_b$, vs. their mistagging efficiency, which it is useful to divide into $c$-mistagging, $\epsilon_c$, and light-jet-mistagging $\epsilon_j$. These efficiencies describe the percentage of true $b$-quark (or $c$-quark or light parton) originating jets which are positively tagged by the algorithm. The efficiencies are generally a function of a cut parameter in the tagging algorithm which can be adjusted to favor greater purity or greater inclusiveness in the tagging.

We can exclude much of the $b\overline{b}$ background by instituting a veto on events with one or more $b$-tags. Based on simulations of ILC tagging efficiency, we take $\epsilon_b = 0.7$, $\epsilon_c = 0.1$ and $\epsilon_j = 0.005$ as a plausible working point. Then for a $b\overline{b}$ background resolved to two jets only $10 \%$ of events will pass the $b$-tag veto. This also applies to roughly a fifth of the large $ee \rightarrow qq$ background. (Backgrounds from $c\overline{c}$ will be reduced by a factor of $\sim 20 \%$ as well.) Thus, although the Higgs decays are now being added to the background, the total number of expected background events can be slightly reduced. For the signal we would expect a small reduction in expected events, mostly due to $W$s decaying to $c$ and $s$ quarks. In the semi-leptonic analysis this would lead to about $5\%$ reduction of the signal. For the all-hadronic analysis, estimation of tagging efficiencies is somewhat ambiguous since we have multiple jets
arising from a single $b$ parton. If we treat each of the four jets according to the efficiencies above, with all jets arising from the
$b\overline{b}$ backgrounds having a ``true'' identity as a $b$-jet, then it is advantageous to veto events with more than one $b$-tag while keeping those with up to one $b$-tag. This would preserve virtually all of the signal while reducing the $b\overline{b}$ backgrounds by $\sim 92 \%$.

The net result of adding $h \rightarrow b\overline{b}$ decays and $b$-tagging is thus a very small change to the expected cross-section sensitivity, on the order of $1-2 \%$ correction.

\bibliographystyle{JHEP}
\bibliography{references}


\end{document}